\documentclass{JHEP3}
\newlength{\vshift}
\newlength{\hshift}
\usepackage{amssymb,amsopn}

\def\la{\lambda}
\def\La{\Lambda}

\def\de{\delta}
\def\be{\beta}
\def\al{\alpha}

\def\ds{\stackrel{\star}{,}}
\def\x{\hat x}

\def\p{\partial}

\def\t{\tilde}

\def\lb{\lbrack}
\def\rb{\rbrack}
\def\pat{\partial}

\title{$U(1)$ gauge field theory on $\kappa$-Minkowski  space}

\author{Marija Dimitrijevi\' c \\ Max-Planck-Institut f\"ur Physik,
        F\"ohringer Ring 6, 80805 M\"unchen, Germany, \\
	Arnold Sommerfeld Center for Theoretical Physics,
	Universit\"at M\"unchen, Fakult\"at f\"ur Physik,
	Theresienstr.\ 37, 80333 M\"unchen, Germany,\\
	University of Belgrade, Faculty of Physics,
	Studentski trg 12, 11000 Beograd, Serbia and Montenegro\\
E-mail:\email{dmarija@theorie.physik.uni-muenchen.de}}
\author{Larisa Jonke\\ Rudjer Bo\v skovi\'c Institute, Theoretical Physics 
Division,
        Bijeni\v cka 54, 10002 Zagreb, Croatia \\ E-mail:\email{larisa@irb.hr}}
\author{Lutz M\"oller \\ Max-Planck-Institut f\"ur Physik,
        F\"ohringer Ring 6, 80805 M\"unchen, Germany
	\\ E-mail:\email{lmoeller1@web.de}}

\abstract{
This study of $U(1)$ gauge field theory on the kappa-deformed Minkowski
spacetime extends previous work on gauge field theories on
this type of  noncommutative spacetime.
We discuss in detail the properties of the Seiberg-Witten map and the
resulting effective
action for $U(1)$ gauge theory with fermionic matter
expanded in ordinary fields. 
We construct the conserved 
gauge current, fix part of the ambiguities in the
Seiberg-Witten map and obtain an effective $U(1)$ action invariant
under the action of the undeformed Poincar\'e group.}


\begin{document}

\section{Introduction}

In  previous papers \cite{f1,f2}  (gauge) field theories
 on the $\kappa$-Minkowski  spacetime were
constructed. All techniques necessary for such
a construction were thoroughly discussed there.
In this paper we concentrate on the $U(1)$ gauge field theory on
 the $\kappa$-Minkowski spacetime. In the first order of expansion in 
 a deformation parameter, we construct an effective $U(1)$ gauge
 field theory, represented on
commutative spacetime, using the $\star$-product
formulation and the Seiberg-Witten map. 
We also construct a conserved gauge 
current.
The appearance  of the additional conserved  current forces us to analyse  
ambiguities in the Seiberg-Witten map.
A specific choice of the Seiberg-Witten map provides 
an effective $U(1)$ action invariant
under the action of the undeformed Poincar\'e group.

The paper is organized as follows:
In the second section we briefly
review some of the properties of the $\kappa$-Minkowski
spacetime. In the third section the
Seiberg-Witten map for gauge fields is constructed and using this result the field strength tensor is calculated.
In the fourth section
we construct the action for $U(1)$ gauge theory with fermionic
matter and analyse the effective
action obtained by expansion in the deformation parameter.
Finally, in the fifth section we construct the conserved gauge current, 
discuss ambiguities of the Seiberg-Witten map  and
construct an  action invariant
under the action of the undeformed Poincar\'e group.

\section{$\kappa$-Minkowski spacetime}

Algebraically, the $n+1$-dimensional $\kappa$-Minkowski spacetime can be introduced~\cite{f1,f2}
as a factor space
of the algebra freely generated by coordinates $\hat{x}^\mu$ divided by
the ideal generated by the following commutation relations:
\begin{equation}
\label{2.1}
\lb \x^\mu , \x^\nu \rb = i C^{\mu\nu}_\rho \x^\rho,
\end{equation}
where
\begin{equation}
\label{2.2}
 C^{\mu\nu}_\rho=a(\delta^\mu _n \delta^\nu_\rho-\delta^\nu_n\delta^\mu_\rho),\quad \mu=0,\dots,n,
\end{equation}
and the (formal) metric of the $\kappa$-Minkowski spacetime is $\eta^{\mu\nu}=diag(1,-1,\ldots,-1)$.
A constant deformation vector $a^\mu$ of length $a$
points to the $n$-th spacelike direction, $a^n=a$, and is related to the 
frequently used parameter $\kappa$ as $a=1/\kappa$. Latin indices denote undeformed  dimensions, $n$ is the deformed
dimension and the Greek indices refer to all $n+1$ dimensions.

There exists an isomorphism between this abstract algebra and 
the algebra of functions of commuting variables equipped with a
$\star$-product \cite{Abe}. Our goal is to construct an effective
field theory on the ordinary spacetime. Therefore we work
in the $\star$-product formalism. The  symmetric $\star$-product for the $\kappa$-Minkowski
spacetime\footnote{There is a standard $\star$-product for Lie algebras
\cite{Kathotia}.},
up to first order in the deformation parameter, is given by
\begin{eqnarray}
\label{sy3}
f(x)\star g(x) &=&
f(x) g(x) + \frac{i}{2}C^{\mu\nu}_\lambda x^\lambda \p_\mu f(x) \p_\nu g(x) \nonumber\\
&=& f(x) g(x) + \frac{ia}{2}x^j \big( \p_n f(x)\p_j g(x) - \p_jf(x) \p_n  g(x)\big) \ .
\end{eqnarray}
From (\ref{sy3}) we have
\begin{equation}
\label{2.33a}
[x^n \ds x^j] = x^n\star x^j -
x^j \star x^n = ia x^j,\;\; [x^i \ds x^j] =0; \;\; i,j=0,1,\ldots,n-1.
\end{equation}

In order to construct derivatives on this space,
we impose several formal conditions that these  derivatives should  fulfil.
The two most critical requirements are first, that the derivatives 
should be consistent
with the commutation relations (\ref{2.33a}) \cite{Wess}, and second,
that they should be
antihermitean  under the integral which we  later introduce to
define the effective action.
The first requirement, together with the demand that derivatives should 
commute among themselves, does not
fix the derivatives sufficiently \cite{forms}. If we add an additional 
requirement that derivatives should
have a vector-like transformation law under  $\kappa$-deformed Lorentz
transformations\footnote{In this paper we omit the representation of the
generators of $\kappa$-Lorentz transformation because they are not
essential for the problem at hand. Note that the $\kappa$-deformed Poincar\' e 
algebra was first introduced in Ref.\cite{Luk}.}, we obtain an almost \cite{forms} 
unique solution
which we call Dirac derivative
\begin{eqnarray}
\label{m1}
\lbrack D^*_n \ds x^j\rbrack &=&-iaD^{j*} , \nonumber \\
\lbrack D^*_n \ds x^n\rbrack &=&\sqrt{1+a^2 D^{\mu *}D^*_\mu} , \nonumber \\
\lbrack D_i^* \ds x^j\rbrack &=&\eta _i^{\ j}\left(
-iaD^*_n+\sqrt{1+a^2D^{\mu *}D^*_\mu}\right) ,\nonumber\\
\lbrack D^*_i \ds x^n\rbrack &=&0 .
\end{eqnarray}
The particular
choice of derivative we use~\cite{f1,forms} is convenient since in this
basis the $\kappa$-Poincare algebra remains undeformed and the
deformation is in the coalgebra sector.
This basis was also used in \cite{k-gn}, where it was called
"classical" basis.
Using  relations (\ref{m1}) we calculate the representation of the derivatives 
$D^*_\mu$  on ordinary functions,
\begin{eqnarray}
D^*_n f(x) &=& \left( {1\over a} \sin (a\partial _n) -
{\cos (a\partial_n) - 1\over ia\partial_n^2}\partial _j\partial ^j\right) 
f(x),\nonumber\\
D^*_i f(x) &=& {e^{-ia\partial _n}-1 \over -ia\partial _n}\partial _i f(x)  .\label{m2}
\end{eqnarray}
Formulae (\ref{m2}) deliver the representation of $D_\mu^*$ derivatives 
in terms of the usual partial derivatives $\partial _\mu$.

With this choice of derivatives the first condition  is fulfilled. 
In order to fulfil the second one as well,
we need $\tilde{D}_\mu^*$ derivatives that are antihermitean under the
integral with a measure $\mu$
\begin{equation}
\label{6.15}
\int d^{n+1} x\hspace{1mm} \mu(x)\hspace{1mm}\bar{f}\star
\tilde {D}^* _\alpha g =-  \int d^{n+1} x\hspace{1mm}
\mu(x)\hspace{1mm}\overline{\tilde{D}^*_\alpha f}\star g .
\end{equation}
This demand will become clearer in section 4.
The derivatives $\tilde{D} _\alpha^*$
are obtained by substituting
$$\p_i \longrightarrow \t{\p _i}=\p_i+\frac{\partial_i\mu}{2\mu},\;\qquad
\p_n\longrightarrow\t{\p _n}=\p_n$$
in (\ref{m2}). The representation of these improved  derivatives on functions is
\begin{eqnarray}
\tilde{D}^*_n f(x) &=& \left( {1\over a} \sin (a\p _n) -
        {\cos (a\pat_n) - 1\over ia\pat_n^2}\tilde{\p} _j\tilde{\p} ^j \right) f(x),\nonumber\\
\tilde{D}^*_i f(x) &=& {e^{-ia\p _n}-1 \over -ia\p _n}\tilde{\p} _i f(x) . \label{k18}
\end{eqnarray}
The Leibniz rules for 
$\tilde{D}_\mu^*$ derivatives (as well as for the $D^*_\mu$
derivatives) are non-trivial as a consequence of (\ref{m1})
\begin{eqnarray}
\tilde{D}^*_n \left( f(x)\star g(x) \right) &=& (D_n^* f(x)) \star
        (e^{-ia \pat_n} g(x)) + (e^{ia\pat_n} f(x)) \star
        (\tilde{D}_n^* g(x))\nonumber\\
&& - ia\left( D^*_je^{ia\pat_n} f(x) \right) \star
        (\tilde{D}^{j*} g(x)), \nonumber\\
\tilde{D}_i^* \left(  f(x)\star g(x) \right) &=& ( D_i^* f(x)) \star
        (e^{-ia\pat_n^*} g(x)) + f(x) \star (\tilde{D}_i^* g(x)).\label{k19}
\end{eqnarray}

In Ref.\cite{forms} it has been shown that the fundamental object to construct 
geometrical theories on $\kappa$-deformed space-times is the 
exterior differential d, which is nilpotent and 
has an undeformed Leibniz rule.
Furthermore, it has been shown that d 
can be decomposed as d$= \xi^{*\mu}D^*_\mu $ 
using a set of n+1 one-forms $\xi^{*\mu}$ 
dual to the Dirac derivative. 
These one-forms (and  all forms in the deRham complex)
are derivative valued, i.e. they have non-trivial commutation 
relations with coordinates, as a consequence of demanding that there are 
only $n+1$ one-forms in $n+1$-dimensional $\kappa$-deformed space-time. 
In the present paper, we go beyond the framework of Ref.\cite{forms}
in requiring from the start the anti-hermiticity of Dirac derivatives 
under an integral. The shift from $\partial_j$ to  $\tilde{\partial}_j$ does 
not affect the conclusions of Ref.\cite{forms}, there exists a set of 
one-forms $\tilde{\xi}^{*\mu}$ dual to the shifted Dirac derivatives 
$\tilde{D}^*_\mu$ such that the exterior differential 
d$=\tilde{\xi}^{*\mu}\tilde{D}^*_\mu $ is nilpotent with an 
undeformed Leibniz rule. 

\section{Gauge theories and the Seiberg-Witten map}

Gauge theories on the $\kappa$-Minkowski spacetime have two new properties, as a
consequence of noncommutativity, see \cite{f2} for details.
A gauge field is both
enveloping algebra-valued and derivative-valued.
Having an enveloping algebra-valued gauge field leads to a theory with infinitely
many degrees of freedom. The way out of this 
unphysical situation is provided in terms of the
Seiberg-Witten map \cite{SW}. 
Using this map one can express noncommutative variables (gauge parameter,
fields) in terms of commutative ones and in 
this way retain the same number of
degrees of freedom as in the commutative case (where the degrees of freedom are Lie algebra-valued).

Explicit solutions of the SW map for the gauge parameter and the matter field are constructed\footnote{In
this section we construct solutions for the non-abelian gauge theory, although starting form the next section
we use solutions for $U(1)$ gauge theory only.}
from the assumption that
\begin{equation}
\label{5}
\de_\al \psi = i \La_\al(x) \star \psi
\end{equation}
and that this is a gauge transformation
\begin{equation}
\label{6}
(\de_\al\de_\be -\de_\be\de_\al)\psi = \de_{\al\times\be} \psi .
\end{equation}
Up to first order in $a$, the solutions\footnote{The Lie algebra-valued field
$A_\mu^0$ has the usual transformation property
$\delta_\alpha A_\mu^0=\partial_\mu\alpha-i[A_\mu^0,\alpha],$
and  the  matter field $\psi^0$ transforms as
$\delta_\alpha \psi^0=i\alpha\psi^0$.}
are
\cite{f2}
\begin{eqnarray}
\Lambda_\al &=&
\alpha -\frac{1}{4}C^{\rho\sigma}_\lambda x^\lambda \{A^0_\rho,\partial _\sigma\alpha\},\label{s1}\\
\psi &=&
\psi^0 -{1\over 2}C^{\rho\sigma}_\la x^\la A^0_\rho\partial _\sigma\psi ^0
+{i\over 8} C^{\rho\sigma}_\la x^\la \lb A^0_\rho,A^0_\sigma\rb \psi ^0 \label{ps2}.
\end{eqnarray}
In the following we  concentrate on the derivative-valued gauge fields.
Since we are using the  modified derivatives $\t{D}^*_\mu$, we have to
modify the solutions of the Seiberg-Witten map for the gauge field $V_\mu$
given in \cite{f2} as well.
The covariant derivative $\t{\cal{D}}_\mu=\t{D}^*_\mu -i\t{V}_\mu$ is defined by its
transformation law
\begin{equation}
\de_\al \Big(\t{\cal{D}}_\mu\psi(x)\Big)
=i{\La}_\al (x) \star \t{\cal{D}}_\mu \psi(x) .\label{s3}
\end{equation}
From (\ref{s3}), it  follows that 
the trasformation law for the gauge field $\t{V}_\mu$ is
\begin{equation}
\delta_\alpha \t{V}_\mu\star \psi = \t{D}^*_\mu \left( \Lambda _\al\star\psi\right)
- \Lambda _\al\star \left(\t{D}^*_\mu\psi\right) +i\Lambda _\al\star\t{V}_\mu\star\psi
-i\t{V}_\mu\star \left(
\Lambda _\al\star\psi\right) .\label{s4}
\end{equation}
Since $\tilde{D}_\mu^*$ derivatives have non-trivial Leibniz rules (\ref{k19}),
we have to treat
the first term on the right-hand side with much care. 
It is convenient to separate
the $n$-th and the $i$-th components of (\ref{s4}).

First, we look at the $i$-th component. Using the Leibniz rule for $\t{D}_i^*$ (\ref{k19}),
we obtain
\begin{equation}
\de_{\al}\t{V}_i\star\psi=(D^*_i \La_{\al})\star\left( e^{-ia\p _n}\psi\right)
+i\Lambda _\al\star\t{V}_i\star\psi-i\t{V}_i\star \left(
\Lambda _\al\star\psi\right) .\label{s6}
\end{equation}
In order to solve this equation, 
we allow derivative-valued gauge field components
$\t{V}_i$ as we did in \cite{f2}. Inserting the ansatz
\begin{equation}
\t{V}_i=V_i\star e^{-ia\p _n} \label{s7}
\end{equation}
in (\ref{s6}) leads
(after  using $e^{-ia\p _n}(f\star g)=(e^{-ia\p _n}f)\star e^{-ia\p _n}g$ and
omitting $e^{-ia\p _n}\psi$ on the right-hand side) to
\begin{equation}
\de_{\al}V_i=(D^*_i \La_{\al})+i\La_{\al}\star V_i -iV _i\star \big( e^{-ia\p _n}\La_{\al}\big) . \label{s8}
\end{equation}
We solve this equation perturbatively, i.e., we expand the $V_i$ field
in the deformation parameter $a$:
\begin{equation}
V_i=V_i^0+V_i^1+\dots  \nonumber
\end{equation}
and use the solution for the gauge parameter $\Lambda _\al$ (\ref{s1}). Up to
first order the solution for $\tilde{V}_i$
is\footnote{The field strength $ F^0_{\mu\nu}$
is the usual field strength of the
undeformed theory;
$F^0_{\mu\nu}=\partial_\mu A^0_\nu - \partial_\nu A^0_\mu-i[A^0_\mu,A^0_\nu].$}
\begin{equation}
\t{V}_i = A^0_i-iaA^0_i\partial _n -{ia\over 2}\partial _n A^0_i-{a\over4}\{A^0_n,A^0_i\}
+{1\over 4} C^{\rho\sigma}_\la x^\la \Big( \{ F^0_{\rho i},A^0_\sigma\} -\{ A^0_\rho ,\partial _\sigma A^0_i\}\Big) .
\label{s9}
\end{equation}
One notices that this solution is the same as in \cite{f2}.
This is to  be expected, because
the $\t{V}_i$ field is only $\p _n$ derivative-valued and
$\p _n$ is not modified.

Next, we  look at the $n$-th component of equation (\ref{s4}). Using the
Leibniz rule for $\t{D}_n^*$ (\ref{k19})
leads to
\begin{eqnarray}
\de_{\al}\t{V}_n\star\psi &=& (D^*_n \La_{\al})\star e^{-ia\partial_n}\psi+
\Big((e^{ia\partial_n}-1) \La_{\al}\Big)\star \t{D}^*_n\psi \nonumber \\
&& -ia(D^*_je^{ia\partial_n}\La_{\al})
\star \t{D}^{j*}\psi  +i\Lambda _\al\star\t{V}_n\star\psi-i\t{V}_n\star \left(
\Lambda _\al\star\psi\right) .\label{s11}
\end{eqnarray}
We make the following ansatz:
\begin{equation}
\t{V}_n=V_{n1}\star e^{-ia\p _n}+V_{2}\star \t{D}^*_n+V_{n3}^l\star \t{D}^*_l \label{s12}
\end{equation}
and insert it in equation (\ref{s11}). Collecting terms proportional to
$\star e^{-ia\partial_n}\psi$, $\star \t{D}^*_n\psi$ and $\star \t{D}_l^*\psi$,
we obtain transformation laws for
the field components $V_{n1}$,
$V_{2}$ and $V_{n3}^l$, respectively:
\begin{eqnarray}
\de_{\al}V_{n1} &=& (D_n^*\Lambda _\al) +i\Lambda _\al\star V_{n1}-iV_{n1}\star (e^{-ia\p _n}\Lambda _\al )
\nonumber \\
&& -iV_{2}\star (D_n^*\Lambda _\al )-iV_{n3}^l\star (D ^*_l\Lambda _\al) ,\label{s13}\\
\de _\al V_{2} &=& \big( (e^{ia\p _n}-1)\Lambda _\al\big) +i\Lambda _\al\star V_{2}
-iV_{2}\star \big( e^{ia\p _n}\Lambda _\al\big) ,\label{s14}\\
\de _\al V_{n3}^l &=& -ia\big( \p ^{l*}\Lambda _\al\big) +i\Lambda _\al\star V_{n3}^l
-iV_{n3}^l\star\Lambda _\al
-aV_{2}\star (\p ^{l*}\Lambda _\al ). \label{s15}
\end{eqnarray}
Up to first order in $a$ the solutions for these equations are
\begin{eqnarray}
V_{n1} &=& A_n^0-{a\over 2}\Big( i\p _jA^{0j}+A_j^0A^{0j}\Big)
+{1\over 4} C^{\rho\sigma}_\la x^\la \Big( \{ F^0_{\rho n},A^0_\sigma\}
-\{ A^0_\rho ,\partial _\sigma A^0_n\}\Big) , \nonumber\\
V_{2} &=& iaA_n^0, \nonumber\\
V_{n3}^j &=& -iaA^{0j}, \nonumber
\end{eqnarray}
and
\begin{eqnarray}
\t{V}_n &=& A^0_n-iaA^{0j}\t{\p} _j -{ia\over 2}\partial _jA^{0j}-{a\over 2}A^0_jA^{0j}\nonumber\\
&&+{1\over 4} C^{\rho\sigma}_\la x^\la \Big( \{ F^0_{\rho n},A^0_\sigma\}
-\{ A^0_\rho ,\partial _\sigma A^0_n\}\Big) .\label{s16}
\end{eqnarray}
Comparing this solution with the solution for $V_n$ in \cite{f2}, we see that
the only difference is
the term $-iaA^{0j}\t{\p} _j$ as a consequence of modifying the derivatives. 

As the next step we construct the field-strength tensor. It
is defined as
\begin{equation}
{\cal F}_{\mu\nu}=i\lb \t{\cal D}_\mu\ds\t{\cal D}_\nu\rb .\label{s17}
\end{equation}
Applying this to the field $\psi$ gives
\begin{eqnarray}
{\cal F}_{ij} &=& \Big( (D_i^*V_j)-(D_j^*V_i)-iV_i\star(e^{-ia\p _n}V_j)+iV_j\star(e^{-ia\p _n}V_i)
\Big) \star e^{-2ia\p _n}, \label{s18}\\
{\cal F}_{nj} &=& F_{nj1}\star e^{-2ia\p _n} +F_{nj2}\star e^{-ia\p _n}\t{D}^*_n
+F_{nj3}^l\star e^{-ia\p _n}\t{D}_l^* , \label{s19}
\end{eqnarray}
where
\begin{eqnarray}
F_{nj1} &=& (D_n^*V_j)-(D_j^*V_{n1})-iV_{n1}\star(e^{-ia\p _n}V_j)+iV_j\star(e^{-ia\p _n}V_{n1})
\nonumber\\
&-&iV_{n3}^l\star (D_l^*V_j)-iV_{n2}\star(D_n^*V_j), \nonumber\\
F_{nj2} &=& \big( (e^{ia\p _n}-1)\big) V_j -(D_j^*V_{n2})-iV_{n2}\star(e^{ia\p _n}V_j)
+iV_j\star(e^{-ia\p _n}V_{n2}),\nonumber \\
F_{nj3}^l &=& -ia(\p ^{l*}V_j) -(D_j^*V_{n3}^l) -aV_{n2}\star (\p ^{l*}V_j)
-iV_{n3}^l\star V_j +V_j\star (e^{-ia\p _n}V_{n3}^l) .\nonumber
\end{eqnarray}

It is obvious that the ${\cal F}_{\mu\nu}$ tensor is derivative valued.
Therefore, we use the same
procedure as in \cite{f2}, namely, we split ${\cal F}_{\mu\nu}$ into the curvature-like terms and
torsion-like terms\footnote{Note that torsion-like terms are defined as the coefficients
 of modified covariant derivatives.}
\begin{equation}
{\cal{F}}_{\mu \nu}=F_{\mu\nu}+T^\rho_{\mu\nu}\t{\cal{D}}_\rho+ \dots +
T^{\rho _1\dots\rho _l}_{\mu\nu}:\t{\cal{D}}_{\rho _1}\dots\t{\cal{D}}_{\rho _l}:+\dots \ .\label{s20}
\end{equation}
Expanding (\ref{s18}) and (\ref{s19}) up to first order in $a$
and rewriting them in this form gives
\begin{eqnarray}
F_{ij} &=& F_{ij}^0 -ia{\cal D}_n^0F_{ij}^0
+{1\over 4} C^{\rho\sigma}_\la x^\la \Big( 2\{F^0_{\rho i},F_{\sigma j}^0\}
+\{ {\cal D}^0_\rho F_{ij}^0,A^0_\sigma \} -\{ A^0_\rho,\partial _\sigma F_{ij}^0 \}\Big) , \nonumber \\
T_{ij}^\mu &=& -2ia\delta ^\mu _nF_{ij}^0, \nonumber\\
F_{nj} &=& F_{nj}^0 -{ia\over 2}{\cal D}^{\mu 0}F_{\mu j}^0 \nonumber\\
&& \quad\>\;\>+{1\over 4} C^{\rho\sigma}_\la x^\la\Big( 2\{F^0_{\rho n},F_{\sigma j}^0\}
+\{ {\cal D}^0_\rho F_{nj}^0,A^0_\sigma \} -\{ A^0_\rho,\partial _\sigma F_{nj}^0 \}\Big) , \nonumber \\
T_{nj}^\mu &=& -ia\eta ^{\mu l}F_{lj}^0-ia\delta ^\mu _nF^0_{nj}. \label{s24}
\end{eqnarray}

These results are the same as in Ref.\cite{f2}. Actually, we have 
checked this up to
second order in $a$, and because of the structure of equations (\ref{s18}),
(\ref{s19}) and (\ref{k19}), we expect that this result
holds to all orders in $a$. In the action for the gauge field,
only curvature-like terms are used
and from equations (\ref{s24}) we see that the modification of the derivatives
does not affect the action.

In this paper we are interested
in a {\it constructive} approach to formulating gauge theories.
Due to this constructive approach we cannot provide a decisive answer
whether the gauge theory presented here can also be formulated geometrically,
with the gauge fields  as components of a
one-form or connection. However, we strongly assume that such geometrical 
formulation is not only possible but is the "proper" formulation of such a
noncommutative gauge theory.

\section{Action}

In order to construct the action, we need an integral with the trace property. This is essential
both for the formulation of the variational principle and
for the gauge invariance of the action.
We  defined such an integral in the $\star$-product formalism, see Ref.\cite{f1}.
There we  used the usual definition of an integral
of functions of commuting variables and introduced
a measure function
to implement the trace property:
\begin{equation}
\label{cyc}
\int d^{n+1}x\; \mu(x)(f\star g) =\int d^{n+1}x\; \mu(x)(g\star f).
\end{equation}
Note that $\mu(x)$ is not $\star$-multiplied with the other
functions, it is a part of the volume element.
From (\ref{cyc}), it follows that
\begin{equation}
\partial_n \mu(x) =0, \qquad x^j \partial_j \mu (x)= -n\mu(x) .\label{mes}
\end{equation}
The measure function is $x^n$ independent and  does not depend on
the deformation parameter $a$ either.
In addition, the measure function is singular at zero and is not 
unique \cite{lt}.
However, after defining the Lagrangian
density in such a way that it vanishes at zero, we can
choose a positive-definite measure function.
Note also that
the explicit form of $\mu(x)$
is not required in any of the subsequent calculations.
We only use relations (\ref{mes}), and therefore non-uniqueness 
of the solution for $\mu(x)$
does not affect our results\footnote{An alternative way of
constructing the  measure function is using a map
which connects the $\kappa$-Minkowski spacetime coordinates and
the spacetime coordinates of the canonical noncommutative
spacetime \cite{am}.}.

With this integral we define the action as follows:
$$S=\int d^{n+1}x\; \mu(x) {\cal L},$$
where  ${\cal L}$ is the Lagrangian density.
Since we saw that the measure function $\mu(x)$ does not vanish in
the limit $a\to 0$, we have to define the Lagrangian density such that
$$\lim_{a\to 0}\mu(x) {\cal L} ={\cal L}^0.$$
Here ${\cal L}$ is the effective Lagrangian density
expanded in powers of the deformation parameter $a$, and
${\cal L}^0$ is the Lagrangian density of the corresponding undeformed field theory.
Although this construction may appear rather arbitrary, we find that
imposing such a "good" limit $a\to 0$ is an important physical requirement.

To this end, we now concentrate on $U(1)$ gauge theory coupled to fermions.
First, we analyse the action for matter fields. It should be the gauge covariant version
of the action for the fermion matter fields defined in Ref.\cite{f1}, so the first guess
would be
\begin{equation}
\label{s0}
S_m=\int d^{n+1} x \;\mu(x) \left(\tilde{\bar{\psi}}\star(i\gamma^{\mu}
\tilde{D}^*_{\mu} +\gamma^{\mu}\t{V}_{\mu}\star -m){\tilde\psi}\right).
\end{equation}
We have chosen the symbol $\tilde{\psi}$ instead of $\psi$ for later convenience.
Using the variational principle
\begin{equation}
\label{var}
\frac{\delta}{\delta g(x)}\int d^{n+1} x\hspace{1mm}
\mu(x)\hspace{1mm} f\star g \star h = \frac{\delta}{\delta g(x)}\int d^{n+1}
x\hspace{1mm}
\mu(x)\hspace{1mm}  g  (h \star f) = \mu(x)\hspace{1mm} (h\star f)
\end{equation}
from the action (\ref{s0}), we obtain
the equation of motion for the matter field $\tilde{\psi}$:
\begin{equation}
\label{e1}
\mu(x)(i\gamma^{\mu}\tilde{\cal D}^*_{\mu} -m)\t{\psi}=0.
\end{equation}
It is obvious that this equation does not have the proper classical limit. In Ref.\cite{f1} this problem was
solved by rescaling the field $\tilde{\psi}$
\begin{equation}
{\tilde\psi}\to \mu^{-1/2}\psi .\label{m3}
\end{equation}
Unfortunately, this rescaling is not fully compatible with the
Seiberg-Witten map. Namely, if $\de_\al \psi = i \La_\al \star \psi$,, then
$$
\delta_\alpha \tilde{\psi} = \delta_\alpha (\mu^{-1/2}\psi) =
i\mu^{-1/2}(\Lambda_\alpha \star\psi)\neq i\Lambda_\alpha \star\tilde{\psi}
$$
and the action (\ref{s0}) will not be gauge invariant.


Nevertheless, demanding  
$$\de_\al \tilde{\psi} = i \La_\al \star \tilde{\psi}$$
we can reconstruct the Seiberg-Witten map for the field $\tilde{\psi}$, but this time taking
the solution in the $a\to 0$ limit as 
$\tilde{\psi}^0=\mu^{-1/2}\psi^0$ instead of $\psi^0$. This is allowed
by the transformation law $\de_\al \tilde{\psi}^0 = i \alpha \tilde{\psi}^0$. 
Repeating the same calculation
we find the following solution:
\begin{equation}
\tilde{\psi} = \mu^{-1/2}\psi^0 -\mu^{-1/2}\frac{1}{2} C^{\rho\sigma}_\la x^\la A^0_\rho\partial _\sigma
\psi ^0 -\mu^{-1/2}\frac{na}{4}A_n^0\psi^0. \label{m5}
\end{equation}

Having obtained the way to rescale the field $\tilde{\psi}$ and using 
the solutions for the Seiberg-Witten map,
we write down the equations of motion from (\ref{e1}) up to first order in $a$
\begin{eqnarray}\label{me}
(i\gamma^{\mu}{\cal D}^0_{\mu}-m)\psi^0
-\frac{1}{2}C^{\rho\sigma}_\lambda\gamma _\rho {\cal D}^0_\sigma {\cal D}^{0\lambda} \psi^0
\hspace*{-3mm}&-&\hspace*{-3mm}\frac{i}{2}C^{\rho\sigma}_\lambda x^\lambda \gamma^\mu F_{\mu\rho}^0({\cal D}^0_\sigma\psi^0) \nonumber\\
&&-\frac{i}{4}C^{\rho\sigma}_\sigma \gamma^\mu F_{\mu\rho}^0\psi^0 =0 ,\nonumber \\
-i\overline{{\cal D}^0_{\mu}\psi^0}\gamma^{\mu}-m\bar\psi^0
-\frac{1}{2}C^{\rho\sigma}_\lambda
\overline{{\cal D}^0_\sigma {\cal D}^{0\lambda} \psi^0} \gamma _\rho 
\hspace*{-3mm}&+&\hspace*{-3mm}\frac{i}{2}C^{\rho\sigma}_\lambda x^\lambda \overline{{\cal D}^0_\sigma\psi^0}\gamma^\mu F_{\mu\rho}^0 
\nonumber\\
&&+\frac{i}{4}C^{\rho\sigma}_\sigma \bar{\psi}^0\gamma^\mu F_{\mu\rho}^0 =0 .
\end{eqnarray}

However, we are also interested in 
the effective action for fermions up to first order in $a$.
Let us write the action (\ref{s0}) with all derivatives explicitly, using (\ref{s7}) and (\ref{s12}):
\begin{eqnarray}
\label{s0d}
S_m=\int d^{n+1} x \;\mu(x) \left(\tilde{\bar{\psi}}\star(i\gamma^{\mu}
\tilde{D}^*_{\mu}-m){\tilde\psi}  +\tilde{\bar{\psi}}\star\gamma^{i}V_{i}\star
e^{-ia\p_n}{\tilde\psi}+\tilde{\bar{\psi}}\star\gamma^{n}V_{n 1}\star
e^{-ia\p_n}{\tilde\psi}\right. \nonumber \\
+\left.\tilde{\bar{\psi}}\star \gamma^nV_{n2}\star\tilde{D}^*_n
{\tilde\psi}+\tilde{\bar{\psi}}\star \gamma^nV_{n3}^j\star\tilde{D}^*_j{\tilde\psi}\right).
\end{eqnarray}
Owing to the cyclicity property of the integral (\ref{cyc}), we can omit
one $\star$ in the
above action. Then we rescale the
fermionic fields using  (\ref{m5})
and, finally, we insert the solutions for the Seiberg-Witten map for the 
gauge field
and obtain up to first order in $a$\footnote{The covariant
derivative ${\cal D}^0_{\mu}$ is the usual covariant derivative
for the undeformed $U(1)$ gauge field theory,
${\cal D}^0_{\mu}=\partial_{\mu}-iA_\mu^0$.}
\begin{eqnarray}
\label{s2}
S_m  = \int d^{n+1}x \left\{ {\bar\psi}^0(i\gamma^{\mu}{\cal D}^0_{\mu}-m)\psi^0 -\frac{1}{4} C^{\rho\sigma}_\lambda x^\lambda {\bar\psi}^0F_{\rho\sigma}^0
(i\gamma^{\mu}{\cal D}^0_{\mu}-m)\psi^0 \right.\nonumber\\
\left.-\frac{1}{2}C^{\rho\sigma}_\lambda\bar{\psi}^0 \gamma _\rho 
{\cal D}^0_\sigma {\cal D}^{0\lambda} \psi^0
-\frac{i}{2}C^{\rho\sigma}_\lambda x^\lambda\bar{\psi}^0 \gamma^\mu F_{\mu\rho}^0({\cal D}^0_\sigma\psi^0) 
-\frac{i}{4}C^{\rho\sigma}_\sigma\bar{\psi}^0 \gamma^\mu F_{\mu\rho}^0\psi^0  \right\} . 
\end{eqnarray}
Since the integral in (\ref{s2}) is the usual integral, applying the variational principle to (\ref{s2}) leads
to the usual Euler-Lagrange equation of motion\footnote{Note that in 
(\ref{s2}), the Lagrangian density depends on
the second derivatives of fields as well.}
\begin{equation}
\partial_\mu \partial_\nu\frac{\partial {\cal{L}}}{\partial(\partial_\mu \partial_\nu \psi)}
- \partial_\mu \frac{\partial {\cal{L}}}{\partial(\partial_\mu \psi)}
+ \frac{\partial {\cal{L}}}{\partial \psi} =0. \label{m7}
\end{equation}
Using (\ref{m7}) the equations of motion for the fields $\bar{\psi}^0$ 
and $\psi^0$ follow from (\ref{s2}).
They are the same as in (\ref{me}), so we do not write them again.

Now we look at the action for the gauge field
\begin{equation}
\label{gauge}
S_g=-\frac{1}{4}\int d^{n+1} x \;\mu(x){\rm Tr}\left(X_2\star F_{\mu\nu}\star F^{\mu\nu}\right).
\end{equation}
Here the gauge covariant expression $X_2$
\begin{equation}
\label{ct}
\delta_\alpha X_2=i[\Lambda_\alpha\ds X_2]
\end{equation}
has been introduced in order to obtain the proper limit $a\to 0$ of the equations of motion, see Ref.\cite{hf}.
From (\ref{ct}) we obtain, up to first order in $a$,
\begin{equation}
X_2=(1-anA^0_n)\mu^{-1}. \label{m8}
\end{equation}
Expanding (\ref{gauge}) up to first order in $a$ and
using the solutions for the Seiberg-Witten map, we obtain
the effective action for the gauge field
\begin{eqnarray}
S_g = -\frac{1}{4}\int d^{n+1}x\left\{F^0_{\mu\nu}F^{0\mu\nu}
-\frac{1}{2}C^{\rho\sigma}_\lambda x^\lambda F^{0\mu\nu}F^0_{\mu\nu}F^0_{\rho\sigma}
+2C^{\rho\sigma}_\lambda x^\lambda F^{0\mu\nu}F^0_{\mu\rho}F^0_{\nu\sigma} \right\} . 
\end{eqnarray}

The complete action for $U(1)$ gauge theory coupled with matter is $S=S_m+S_g$.
The equations of motion  for
the matter  fields are given in equations (\ref{me}).
Using the standard Euler-Lagrange equation
of motion, for the gauge field we obtain
\begin{eqnarray}
&&-J^\rho = \partial_\mu F^{0\mu\rho}
-\frac{1}{2}C^{\alpha\beta}_\mu F^{0\mu\rho}F^0_{\alpha\beta}
-\frac{1}{4}C^{\mu\rho}_\mu F^{0\alpha\beta}F^0_{\alpha\beta}
+C^{\alpha\beta}_\mu F^{0\mu}_\alpha F^{0\rho}_\beta \nonumber\\
&&-C^{\alpha\rho}_\mu F^{0\beta\mu}F^0_{\alpha\beta}
+C^{\alpha\mu}_\mu F^{0\beta\rho}F^0_{\alpha\beta} \nonumber\\
&&+C^{\alpha\beta}_\lambda x^\lambda\p_\mu \big( F^{0\mu}_\alpha F^{0\rho}_\beta 
-\frac{1}{2}F^{0\mu\rho}F^0_{\alpha\beta} \big)
-\frac{1}{4}C^{\mu\rho}_\lambda x^\lambda \p_\mu (F^{0\alpha\beta}F^0_{\alpha\beta}) \nonumber \\
&&-C^{\alpha\rho}_\lambda x^\lambda \p_\mu (F^{0\beta\mu}F^0_{\alpha\beta})
+C^{\alpha\mu}_\lambda x^\lambda \p_\mu (F^{0\beta\rho}F^0_{\alpha\beta}) .\label{3} 
\end{eqnarray}
The current $J^{\rho}$ is given by
\begin{eqnarray}
J^\rho&=&-\partial_\mu\frac{\partial{\cal L}_m}{\partial(\partial_\mu A^0_\rho)}
+\frac{\partial{\cal L}_m}{\partial A^0_\rho} \nonumber\\
&=&\bar{\psi}^0\gamma^\rho\psi^0
-\frac{1}{2}C^{\alpha\beta}_\lambda x^\lambda \bar{\psi}^0\gamma^\rho F_{\alpha\beta}^0\psi^0
-C^{\alpha\rho}_\lambda x^\lambda \bar{\psi}^0\gamma^\mu F_{\mu\alpha}^0\psi^0
-\frac{i}{2}C^{\alpha\mu}_\lambda \eta^{\lambda\rho}\overline{{\cal{D}}^0_\mu\psi^0}\gamma_\alpha\psi^0 
\nonumber\\
&&-\frac{i}{2}C^{\alpha\rho}_\mu \bar{\psi}^0\gamma^\mu ({\cal{D}}^0_\alpha\psi^0)
+\frac{i}{2}C^{\alpha\rho}_\mu \bar{\psi}^0\gamma_\alpha ({\cal{D}}^{0\mu}\psi^0)
+\frac{i}{4}C^{\alpha\mu}_\alpha\Big(\overline{{\cal D}^0_\mu\psi^0}
\gamma^\rho\psi^0 -\bar{\psi}^0\gamma^\rho{\cal D}^0_\mu\psi^0\Big) \nonumber\\
&&+\frac{i}{2} C^{\alpha\mu}_\lambda x^\lambda
 \overline{{\cal D}^0_\mu\psi^0}\gamma^\rho({\cal D}^0_\alpha\psi^0) .\label{J4}
\end{eqnarray}

\section{Symmetries and the Seiberg-Witten map}

Using the  equations of motion (\ref{me}) one can  show that
the current (\ref{J4}) is conserved, $\partial_{\mu} J^{\mu}=0$.
In the undeformed gauge theory, existence and conservation of the 
current $J^\mu$  are 
consequences of the symmetry of the action with respect 
to gauge transformations. One  expects
that the same applies here. To check this, 
we calculate the variation of the action (\ref{s2}) when
$\delta_\alpha \psi^0=i\alpha\psi^0$, $\delta_\alpha \bar{\psi}^0 =-i\alpha\bar{\psi}^0$ and
$\delta_\alpha A_\mu^0=\partial_\mu \alpha$\footnote{Note that $\delta_\alpha F_{\mu\nu}^0=0$.}
\begin{equation}
\delta S_m=\int d^{n+1}x \alpha\;\p_\mu j^\mu=0 . \label{m9}
\end{equation}
Here $j^\rho$ is given by
\begin{eqnarray}
j^{\rho} & = & \bar\psi^0\gamma^{\rho}\psi^0
-\frac{1}{4}C^{\alpha\beta}_\lambda x^\lambda \bar{\psi}^0\gamma^\rho F_{\alpha\beta}^0\psi^0
-\frac{1}{2}C^{\alpha\rho}_\lambda x^\lambda \bar{\psi}^0\gamma^\mu F_{\mu\alpha}^0\psi^0 \label{m11}\\
&&-\frac{i}{4}C^{\alpha\mu}_\lambda\eta^{\lambda\rho}\Big(\overline{{\cal D}^0_\mu\psi^0}
\gamma_\alpha\psi^0 -\bar{\psi}^0\gamma_\alpha{\cal D}^0_\mu\psi^0\Big)
-\frac{i}{4}C^{\alpha\rho}_\lambda\Big(\overline{{\cal D}^{0\lambda}\psi^0}
\gamma_\alpha\psi^0 -\bar{\psi}^0\gamma_\alpha{\cal D}^{0\lambda}\psi^0\Big) .\nonumber
\end{eqnarray}
Comparing this result with (\ref{J4}) there seem to be two different conserved 
currents in our theory. A
difference between these two currents is not topological, nor are
the corresponding conserved charges equal.
Apparently, we have additional (gauge) symmetry in the model.
The source of this symmetry must be  the Seiberg-Witten map.

It is well known that the Seiberg-Witten map is not unique~\cite{FB}. 
An analysis of
the ambiguities in the Seiberg-Witten map in the canonical noncommutative space was provided in
Ref.\cite{lutz2} and we adapt this analysis to the problem at hand.
The important difference with the respect to the canonical case
is that we allow derivative-valued gauge fields as solutions
of the Seiberg-Witten map. In our setting  the derivative-valued
gauge fields appear naturally, as a consequence of non-trivial Leibniz rules for the Dirac
operator (\ref{k19}).
Note that we discuss only the ambiguities relevant to the classical action,
compare~\cite{renorm}.

In Ref.\cite{lutz2} it was shown that the possible ambiguities in the Seiberg-Witten map
for the gauge parameter did not affect the action.
On the other hand, the solution for the Seiberg-Witten map for fermions 
(\ref{m5})
allows an additional term,
\begin{equation}
\label{mat}
\Delta\tilde\psi=
b_1\mu^{-1/2}C^{\rho\sigma}_\lambda x^\lambda F_{\rho\sigma}^0\psi^0,
\end{equation}
which does affect the action.

Furthermore, to the solution for the vector fields (\ref{s9}) and (\ref{s16})
we can add the
following terms\footnote{These additional terms are obtained either as a
solution of the homogeneous part of equation (\ref{s6}), or starting from a
more general ansatz for the gauge field (\ref{s7}):
$\tilde V_j=V_{j1}\star e^{-ia\p_n}+
V_{j2}^n\star \t{D}^*_n+V_{3}\star \t{D}^*_j+V_{j4}^l\star \t{D}^*_l$}:
\begin{eqnarray}
\label{dV}
\Delta \t{V}_\mu &=& ib_2C_{\la}^{\rho\sigma}x^{\la}F^0_{\mu\rho}\tilde{\cal D}
^0_\sigma
+ ib_3C_{\la}^{\rho\sigma}x^{\la}F^0_{\rho\sigma}\tilde{\cal D}^0_\mu +
\nonumber\\
&&+ \frac{i}{2}(b_2-2b_3)C^{\rho\sigma}_\lambda x^\lambda ({\cal{D}}_\sigma^0F^0_{\mu\rho})
+\frac{ia}{2}(nb_2-2b_3)F^0_{\mu n},
\end{eqnarray}
where the coefficients of the last two terms are fixed demanding that $\Delta \t{V}_\mu$ should be hermitean.
Note also that ${\cal{D}}_\sigma^0F^0_{\mu\rho}=\partial_\sigma^0F^0_{\mu\rho}$ because we work with $U(1)$
gauge theory.
This results in the modification of the curvature-like terms
\begin{eqnarray}\label{dF}
\Delta F_{\mu\nu} &=& 2b_2C_{\la}^{\rho\sigma}x^{\la}F_{\mu\sigma}F_{\nu\rho}+
2b_3C_{\la}^{\rho\sigma}x^{\la}F_{\mu\nu}F_{\rho\sigma}+\nonumber\\
&&-\frac{i}{2}(b_2-2b_3)C^{\rho\sigma}_\lambda x^\lambda \Big( {\cal{D}}^0_\nu{\cal{D}}^0_\sigma F^0_{\mu\rho}
- {\cal{D}}^0_\mu{\cal{D}}^0_\sigma F^0_{\nu\rho}\Big) -\frac{ia}{2}(n-1)b_2{\cal{D}}^0_n F^0_{\mu\nu}
\end{eqnarray}
and torsion-like terms
\begin{eqnarray}\label{dT}
\Delta T_{\mu\nu}^\sigma &=&-ib_2\Big( C_{\nu}^{\rho\sigma}F^0_{\mu\rho}- C_{\mu}^{\rho\sigma}F^0_{\nu\rho}\Big)
-ib_2C^{\rho\sigma}_\lambda x^\lambda({\cal{D}}^0_\nu F^0_{\mu\rho} -{\cal{D}}^0_\mu F^0_{\nu\rho}) \nonumber \\
&&-ib_3\Big( \delta^\sigma_\mu ( C^{\rho\alpha}_\nu F^0_{\rho\alpha}
+ C^{\rho\alpha}_\lambda x^\lambda ({\cal{D}}^0_\nu F^0_{\rho\alpha}) )
-\delta^\sigma_\nu ( C^{\rho\alpha}_\mu F^0_{\rho\alpha}
+ C^{\rho\alpha}_\lambda x^\lambda ({\cal{D}}^0_\mu F^0_{\rho\alpha}) )\Big) ,
\end{eqnarray}
of the field strength (\ref{s24}).

Finally it is possible to add the following expression 
to the solution (\ref{m8}) for $X_2$:
\begin{equation}
\Delta X_2 = b_4\mu ^{-1}C^{\rho\sigma}_\lambda x^\lambda F^0_{\rho\sigma} . \label{m12}
\end{equation}

Taking into consideration all the additional terms (\ref{mat}), (\ref{dV}),
(\ref{dF}) and (\ref{m12}), we obtain
a more general  effective action:
\begin{eqnarray}
\label{sa12}
S&=&\int d^{n+1}x\left\{{\bar\psi}^0(i\gamma^{\mu}{\cal D}^0_{\mu}-m)\psi^0
-\frac{1}{4}F^0_{\mu\nu}F^{0\mu\nu}\right.\nonumber\\
&& -\frac{1}{4}C^{\rho\sigma}_\lambda
\left( {\bar\psi}^0\gamma_\rho {\cal D}^0_\sigma{\cal D}^{0\lambda}\psi^0
+\overline{{\cal D}^0_\sigma{\cal D}^{0\lambda}\psi^0}\gamma_\rho\psi^0 \right) \nonumber \\
&&-\frac{1}{4}(1-8b_1)C^{\rho\sigma}_\lambda
x^\lambda{\bar\psi}^0 F_{\rho\sigma}^0 (i\gamma^{\mu}({\cal D}^0_{\mu}\psi^0)-m\psi^0)
-\frac{i}{2}(1-2b_2)C^{\rho\sigma}_\lambda x^\lambda {\bar{\psi}}^0\gamma^\mu F^0_{\mu\rho}
({\cal D}^0_{\sigma}\psi^0) \nonumber\\
&& +ib_3C^{\rho\sigma}_\lambda x^\lambda {\bar{\psi}}^0\gamma^\mu F^0_{\rho\sigma}
({\cal D}^0_{\mu}\psi^0)  -2i(b_1-\frac{1}{4}(b_2-2b_3))C^{\rho\sigma}_\lambda x^\lambda
{\bar{\psi}}^0\gamma^\mu ({\cal{D}}^0_\sigma F^0_{\mu\rho})\psi^0 \nonumber\\
&&-\frac{ia}{4}(n + 8b_1 - 2nb_2 + 4b_3- 1){\bar{\psi}}^0\gamma^\mu F^0_{\mu n}\psi ^0 \nonumber\\
&&\left .
-\frac{1}{2}(1-2b_2)C^{\rho\sigma}_\lambda x^\lambda F^{0\mu\nu}F^0_{\mu\rho}F^0_{\nu\sigma}
+\frac{1}{8}(1-8b_3-2b_4)C^{\rho\sigma}_\lambda x^\lambda F^{0\mu\nu}F^0_{\mu\nu}F^0_{\rho\sigma} \right\} .
\end{eqnarray}

All constants $b_i$ are
completely undetermined, and were all set to
zero in previous calculations.
The reason for this particular  choice was a technical simplicity in
constructing
the Seiberg-Witten map. However, we have  another interesting possibility.
There exist a particular choice of the constants $b_i$ such that all
ambiguous, undetermined terms in the action (\ref{sa12}) are set to zero.

For massless fermions, we choose $b_1=1/16,b_2=1/2,b_3=1/8, b_4=0$, 
and for massive fermions, we choose $b_1=1/8,b_2=1/2,b_3=0, b_4=1/2$.
The effective action for $U(1)$ gauge theory with fermionic 
matter\footnote{Of course, $m=0$ in the massless case.} up to
first order in the deformation parameter $a$ is 
\begin{eqnarray}
\label{final}
S&=&\int d^{n+1}x\left\{{\bar\psi}^0(i\gamma^{\mu}{\cal D}^0_{\mu}-m)\psi^0
-\frac{1}{4}F^0_{\mu\nu}F^{0\mu\nu} \right. \nonumber\\
&&- \frac{1}{4}C^{\rho\sigma}_\lambda
\left. \left( {\bar\psi}^0\gamma_\rho {\cal D}^0_\sigma{\cal D}^{0\lambda}\psi^0
+\overline{{\cal D}^0_\sigma{\cal D}^{0\lambda}\psi}^0\gamma_\rho\psi^0 \right)\right\} .
\end{eqnarray}
The corresponding equations of motion are given as
\begin{eqnarray}\label{mea}
&&(i\gamma^{\mu}{\cal D}^0_{\mu}-m)\psi^0
-\frac{1}{2}C^{\rho\sigma}_\lambda\gamma _\rho {\cal D}^0_\sigma {\cal D}^{0\lambda} \psi^0
-\frac{i}{4n}C^{\rho\sigma}_\sigma \gamma^\mu F_{\mu\rho}^0\psi^0 =0, \nonumber \\
&&-i\overline{{\cal D}^0_{\mu}\psi^0}\gamma^{\mu}-m\bar\psi^0
-\frac{1}{2}C^{\rho\sigma}_\lambda
\overline{{\cal D}^0_\sigma {\cal D}^{0\lambda} \psi^0} \gamma _\rho 
+\frac{i}{4n}C^{\rho\sigma}_\sigma \bar{\psi}^0\gamma^\mu F_{\mu\rho}^0 =0, \\
&&\p_\mu F^{0\rho\mu}=J^{\rho}= \bar\psi^0\gamma^{\rho}\psi^0 \nonumber\\
&&-\frac{i}{4}\left( C^{\alpha\mu}_\lambda\eta^{\lambda\rho}\Big(\overline{{\cal D}^0_\mu\psi^0}
\gamma_\alpha\psi^0 -\bar{\psi}^0\gamma_\alpha{\cal D}^0_\mu\psi^0\Big)
+C^{\alpha\rho}_\lambda\Big(\overline{{\cal D}^{0\lambda}\psi^0}
\gamma_\alpha\psi^0 -\bar{\psi}^0\gamma_\alpha{\cal D}^{0\lambda}\psi^0\Big)\right) . \nonumber
\end{eqnarray}
Calculating the current $j^\rho$ from the variation of the action (\ref{m9}) with the above
choice of constants, we obtain $j^\rho= J^\rho$.
Furthermore, we obtain the action (\ref{final}) and equations of motion
(\ref{mea}) that are  both gauge invariant and invariant
 under classical Poincar\'e transformations!

We end this analysis with a few comments. First, note that 
$j^\rho= J^\rho$ means that the fermionic action, up to first order in
the deformation parameter, is
\begin{eqnarray}\label{final2}
S_m = \frac{1}{2}\int d^{n+1}x\left\{
{\bar\psi}^0(i\gamma^{\mu}{\cal D}_{\mu}-m)\psi^0+
(-i\overline{{\cal D}_{\mu}\psi}\gamma^{\mu}-m{\bar\psi})\psi\right\},
\end{eqnarray}
where the operator ${\cal D}_{\mu}$ is the Dirac operator (\ref{m2}) 
expanded in $a$ in 
which partial derivatives are covariantised by the minimal substitution
$\p_\mu\rightarrow \p_\mu-iA^0_\mu$.
We conjecture that this fact might be valid to all orders, but one
needs to be careful in ordering derivatives
in the expansion of the Dirac operator.

Note also that if one allows for the derivative-valued gauge fields in  
the canonical case, one can construct an effective $U(1)$ action with no 
additional terms in the first order with respect to the 
undeformed $U(1)$ gauge theory\footnote{This has already been known to L.M., 
see Ref.\cite{phd}.} (compare with Ref.\cite{jurco}).  
For the classical theory, one can use 
the Seiberg-Witten map (gauge freedom) to transform  
additional gauge interactions 
into the geometry of spacetime (the torsion-like part of the field strength).
Unfortunately, we still do not have a clear understanding of the interplay
between deformed symetries, gauge theories and spacetime geometry.
As a next step, an investigation of deformed general 
coordinate transformations is performed in Ref.\cite{bec}.

\acknowledgments
We thank Frank Meyer and Julius Wess for their
help and many useful discussions.

L. J.  gratefully acknowledges the support of
the Ministry of Science and Technology
of the Republic of Croatia under the contract 0098003 and partial support
of
the Alexander
von Humboldt Foundation.

\end{document}